\documentclass[article,twocolumn,superscriptaddress,nopacs,floatfix,amssymb,amsmath]{revtex4}
\usepackage{epsfig}
\usepackage{graphics}
\usepackage{array}
\usepackage{multirow}
\newcommand{\beqn}{\begin{eqnarray}}
\newcommand{\eeqn}{\end{eqnarray}}

\begin{document}

\title{  Soliton concepts  and the protein structure
}

\author{Andrei Krokhotin}
\email{Andrei.Krokhotine@cern.ch}
\affiliation{Department of Physics and Astronomy, Uppsala University,
P.O. Box 803, S-75108, Uppsala, Sweden}
\author{Antti J. Niemi}
\email{Antti.Niemi@physics.uu.se}
\affiliation{
Laboratoire de Mathematiques et Physique Theorique
CNRS UMR 6083, F\'ed\'eration Denis Poisson, Universit\'e de Tours,
Parc de Grandmont, F37200, Tours, France}
\affiliation{Department of Physics and Astronomy, Uppsala University,
P.O. Box 803, S-75108, Uppsala, Sweden}
\author{Xubiao Peng}
\email{xubiaopeng@gmail.com}
\affiliation{Department of Physics and Astronomy, Uppsala University,
P.O. Box 803, S-75108, Uppsala, Sweden}

\begin{abstract}
Structural classification shows that the number of different protein folds 
is surprisingly small. It also appears that proteins are built in a modular fashion, from 
a relatively small number of  components. Here we propose to identify
the modular building blocks  of proteins with  the dark soliton solution of a generalized 
discrete nonlinear Schr\"odinger equation. For this we  show  that practically all
protein loops  can be obtained simply by scaling  the size and by joining together a 
number of copies of the soliton, one after another. The soliton has only two loop specific parameters and 
we identify their possible values in Protein Data Bank.  
We show that with a collection  of 200 sets 
of parameters, each determining  a soliton profile 
that describes a different short loop, we cover over 90 $\%$ of  all 
proteins with experimental accuracy.   We also present two examples  that describe how
the loop library  can be employed both to model  and to
analyze the structure of folded proteins. 
\end{abstract}

\pacs{ 87.15.A- ,  87.15.Cc}
\maketitle

\section{Introduction}

Proteins come in many shapes, but the number  of different folds is definitely much smaller than
suggested by Levinthal's  estimate \cite{lev}. For example, thus far the structural classification scheme
SCOP \cite{scop} has identified 1393 unique folds while in CATH \cite{cath} here are currently 
1282 topologies. These figures 
have not changed since the year 2008, indicating that the number  of different protein conformations 
is quite limited and probably most of them have already been observed.
Furthermore, the great success of SCOP, CATH and other approaches such as FSSP \cite{fssp} 
in classifying  the architecture of proteins is a manifestation that proteins are built in a modular 
fashion from a relatively small number of different  components. 

Here we advocate a quantitative energy function based approach to identify and classify the 
modular components  of proteins. We propose to utilize  the dark soliton solution of a generalized 
discrete nonlinear  Schr\"odinger (DNLS) equation  as the basic modular building block.
The original  DNLS equation \cite{dnls}, \cite{fadde} shares a long history 
with protein research. The equation was  introduced by Davidov to explain how an energy 
excitation propagates along the $\alpha$-helix \cite{davi}, \cite{scott}. 
The soliton evokes a deformation of the protein shape,  and as a consequence a trapped soliton
is a natural cause for protein folding. The present generalization of the original 
DNLS equation is motivated by recent observations that
protein loops in the HP35 villin headpiece with Protein Data Bank (PDB)  \cite{pdb} code 
1YRF  \cite{nora}, and in the myoglobin with PDB code 1ABS  \cite{peng}
are accurately described in terms of its dark soliton.
In this article  we  extend this observation to essentially all proteins in PDB.   
We  propose to classify the shapes of loops in terms of a small number of universal 
parameters that appear in the generalized DNLS equation.
These parameters specify the global characteristics such as the size and 
location of a short loop that is described by a single soliton. 
But the detailed shape of this loop is entirely determined by the soliton solution. 
Each set of the soliton parameters then corresponds to a different short fundamental loop and these 
fundamental loops constitute the modular building blocks of proteins.

We adopt the present experimental precision  as the quantitative criterion for identifying
two different protein structures.   
The accuracy  of  x-ray measurements which is the dominant approach 
to structure determination, is measured by the  B-factor. 
For very high resolution structures  the  backbone C$_\alpha$ carbons have  B-factor values that  are
typically less than \cite{petsko}
\begin{equation}
B_{max} \ \buildrel < \over \sim \ 
35 \ \dot {\mathrm A}^2
\label{bmax}
\end{equation}
According to the Debye-Waller relation this corresponds to a fluctuation distance that is
less than or equal to
\begin{equation}
\sqrt{ <x^2 >}_{max}  \ = \  \sqrt{ \frac{B_{max}}{8\pi^2}} \ \approx 0.65 { \ \dot {\mathrm A}}
\label{x}
\end{equation}
Consequently 
we identify two structures if they deviate from each other no more than
$ 0.6 - 0.7$ \.A in RMSD. Indeed, when the RMSD value between two loop configurations
is less than this cut-off value, present  experimental techniques  can not  reliably differentiate  
between them so that  for all practical purposes the two structures are identical. 
Here we show  that  it is sufficient to introduce only 200  distinct  parameter sets for the soliton,
constructed using 44 different proteins, 
in order to describe over 90$\%$ of known protein structures with the B-factor 
accuracy.  Consequently the number of different modular protein components appears to be 
almost an order of magnitude smaller than suggested by the present SCOP and CATH data. Since the purpose 
here is to show that we have a method that works, we do not aim to 
optimize  the loop library. But we suspect that  the actual number of truly independent 
loops is  much smaller, probably less than 100. For this we show that the 200 
fundamental loops  can be described  by 57 multiple covered loops.

\vskip 0.3cm
\noindent

\section{Model}

We  characterize the shape of a protein  in terms of  its central C$_\alpha$ backbone.
These carbon atoms  are located at the positions $\mathbf r_i$ where $i=1,...,N$ label the residues. 
For each pair of  nearest neighbors 
${\bf r}_{i+1}$ and ${\bf r}_i$ we introduce the unit tangent vector and the unit bi-normal vector, respectively
\begin{equation}
{\bf t}_i = \frac{ {\bf r}_{i+1} - {\bf r}_i }{ | {\bf r}_{i+1} - {\bf r}_i |}
\hskip 1.0cm \& \hskip 1.0cm {\bf b}_i = \frac{ {\bf t}_{i-1} \times {\bf t}_i }{| {\bf t}_{i-1} \times {\bf t}_i|} 
\label{tb}
\end{equation}
Then 
\begin{equation}
\psi_{ i} = \arccos ( {\bf t}_{i+1} \cdot {\bf t}_i )  \ \ \ \ \& \ \ \ \ 
\theta_{i} = \arccos ({\bf b}_{i+1} \cdot {\bf b}_i)  
\label{bt}
\end{equation}
are the  standard discrete  Frenet frame  bond angle and  torsion angle of the backbone.  
Note that the bond angle $\psi_i$ is determined by three C$_\alpha$ carbons, those
at the sites $\mathbf r_i$, $\mathbf r_{i+1}$ and $\mathbf r_{i+2}$. But for the torsion angle 
$\theta_i$ we need four C$_\alpha$ carbons,
those between sites $i-1$ and $i+2$.  Inversely if the bond and torsion 
angles are known we can reconstruct the entire protein backbone by solving the discrete
Frenet equation. We refer to \cite{dff} for details of the present coordinate system.

An excellent approximation to the standard right-handed $\alpha$-helix  and the $\beta$-strand is  
obtained by setting
\begin{equation}
(\psi_{i}, \theta_i)_\alpha   \ \approx  \  ( \frac{\pi}{2} ,1) \ \ \ \ \ \& \ \ \ \ \
(\psi_{i}, \theta_i)_\beta \   \approx  \ (1,    \pi)
\label{rhab}
\end{equation}
Similarly, we get the other familiar regular
secondary structures like 3/10 helices, left-handed helices {\it etc.} by selecting proper constant values for 
the bond and torsion angles.   We also record that the following $\mathbb Z_2$ 
transformation leaves the backbone coordinates  
intact  \cite{dff}
\begin{equation}
\begin{matrix}
\ \ \ \ \ \ \ \ \ \psi_{k} & \to  &  
 - \ \psi_{k}
 \ \ \ \ \ \ \  
{\rm for \ \ all} \ \  k \geq i  \\
\ \ \ \ \ \ \ \ \ \theta_{i \ }  & \to &  \hskip -2.5cm \theta_{i} - \pi  
\end{matrix}
\label{dsgau}
\end{equation}

Loops are configurations that bridge between these regular secondary structures. 
Elsewhere \cite{nora}, \cite{peng}  it has been shown that loops in the chicken villin headpiece with
PDB code 1YRF and  the myoglobin 1ABS  can  be described in terms
of the dark soliton of the generalized discrete nonlinear Schr\"odinger equation that derives
from the energy function
\[
E = - \sum\limits_{i=1}^{N-1}  2\, \psi_{i+1} \psi_i  
\]
\begin{equation}
+ \sum\limits_{i=1}^N
\biggl\{  2 \psi_i^2 + q\cdot (\psi_i^2 - \mu^2)^2  
 + \frac{r}{2} \cdot \psi_i^2 \theta_i^2  -  v \cdot  \theta_i   +  \frac{w}{2}\cdot  \theta^2_i 
\biggr\} 
\label{E1}
\end{equation}
where $(q,\mu,r,v,w)$ are parameters.
Here the first sum together with the three first terms in the second sum comprise  exactly   the 
energy of  the standard  DNLS equation    \cite{nora}.  The fourth ($v$) is a 
conserved quantity in the DNLS hierarchy  \cite{fadde}, 
called  the "helicity". We note that the conserved "momentum" could also
be added \cite{fadde} but since the improvement in accuracy is minor we leave it out.   
The last ($w$) is the  Proca mass term that we include for completeness.
In this manner the functional form  (\ref{E1}) becomes deeply anchored  in the 
elegant mathematical structure of integrable hierarchies \cite{fadde}. 
But unlike {\it e.g.} force fields in molecular dynamics, the
energy function (\ref{E1}) does not purport to explain the fine 
details of the atomary level mechanisms that give rise to protein folding. Instead, in line with
Landau-Lifschitz  theories it describes the properties of a folded protein 
backbone in terms of universal physical arguments. 

In \cite{nora} it has been shown  that (\ref{E1}) supports 
solitons. For this we first eliminate the variable $\theta_i$
in terms of $\psi_i$,
\begin{equation}
\theta_i [\psi_i] = \frac{v}{ w + r\, \psi_i^2} \equiv \frac{b}{ 1 + e\, \psi_i^2} 
\label{Etau}
\end{equation}
If the value of $\theta_i$ falls outside of its fundamental domain $[-\pi ,\pi]$ 
we redefine it modulo $2\pi$.

We vary the energy function with respect to $\psi_i$ and  
substitute $\theta_i[\psi_i]$ from  (\ref{Etau})  to arrive at
\begin{equation}
\psi_{i+1} - 2 \psi_i + \psi_{i-1} \ = \ U' [\psi_i] \psi_i  \ \equiv\ \frac{dU[\psi]}{d\psi_i^2} \ \psi_i \ \ \ \ (i=1,...,N)
\label{Ekappa}
\end{equation}
with $\psi_{0} = \psi_{N+1} = 0$.  This is a generalization of the DNLS equation with
\[
U[\psi] = -   \frac{1}{2} v \cdot \theta[\psi]   - 
2q \mu^2  \cdot  \psi^2 + 
q \cdot \psi^4
\]
\begin{equation}
=  - (2 q \mu^2 - \frac{1}{2} v b e) \psi^2 + ( q - \frac{1}{2} vbe^2) \psi^4 + ...
\label{U}
\end{equation} 
where we recognize the familiar structure of  the nonlinear 
Schr\"odinger equation potential \cite{dnls}-\cite{scott}. Indeed, it turns out that in the case
of proteins the correction terms give rise to an adjustment that is tiny in comparison to the B-factor accuracy. 

The exact dark soliton solution to the discrete nonlinear Schr\"odinger equation is not known 
in a closed form. But it should be a discrete version  of the continuum solution, and thus
an excellent approximation is obtained by {\it naive} discretization of the 
continuum dark NLSE soliton  \cite{dnls}-\cite{scott}:
\begin{equation}
\psi_i  =  \frac{ 
(m_{1} + 2\pi N_1)  \cdot e^{ c_{1} ( i-s)  } - (m_{2} + 2\pi N_2) \cdot e^{ - c_{2} ( i-s)}  }
{e^{ c_{1} ( i-s) } +  e^{ - c_{2} ( i-s)}   }
\label{bond}
\end{equation}
Here $s$ is a parameter that determines the backbone site location of the center of the fundamental loop 
that is described by 
the soliton.  The $m_{1,2} \in [0,\pi] $ are parameters  that in the continuum limit coincide known combinations
of the parameters in (\ref{U}) \cite{dnls}-\cite{scott}; in the case of proteins their
values  are entirely determined by the adjacent helices and strands. The $N_1$ and $N_2$ constitute the integer parts 
of $m_{1,2}$, initially we take $N_1 = N_2 \equiv N$. This integer is like a covering number, it determines 
how many times $\psi_i$ covers  its fundamental domain $[-\pi, \pi]$  when we traverse the loop once.
Negative values of $\psi_i$ are related to the positive values by (\ref{dsgau}).
Notice that for $m_1 = m_2$  and $c_1 = c_2$ we  recover the hyperbolic tangent. 
Moreover,  {\it only}  the  $c_1$ and  $c_2$ are 
intrinsically loop specific parameters, they specify the length of the loop and as in the case of the $m_{1,2}$, in the
continuum limit they are known combinations of the parameters in (\ref{U}).
Whenever  $\psi_i$ takes values outside of the fundamental domain $[-\pi, \pi]$, we
redefine it modulo $2\pi$. 

A full protein chain is the sum of terms of the form (\ref{bond}), over all the locations of  the centers of its fundamental loops.

As a parameter basis for the soliton description of loops, we use the parameters in (\ref{bond}), (\ref{Ekappa}):
After determining the values of the parameters in (\ref{bond}), we compute the torsion $\theta_i$ from (\ref{Etau})
and construct the curve using the discrete Frenet equation. Notice that since there are only two independent parameters $b$ and $e$ for each fundamental loop in (\ref{Etau}), they are both specified by the regular secondary 
structures that are adjacent to this  loop.  {\it All}  intrinsic loop dependence is  due to $\psi_i$.  

Since our aim is to describe protein structures entirely in terms of
DNLS solitons, hereafter we always define helices and strands and other 
similar regular secondary structures strictly in terms of their {\it geometry}, by using  
($\psi,\theta$) values such as (\ref{rhab}).
The fundamental loops,   the helices and  the strands are then all on similar conceptual footing in the sense that
each of these structures are specified  by two parameters. 
In particular, a fundamental loop  coalesces into a helix or a strand at exponential rate,
when the distance $|i-s|$  from its center increases.

\section{Parameter determination}

The challenge we now need to address is to  enumerate the possible values of the parameters 
 (\ref{bond}) and (\ref{Etau}) in case of PDB proteins.
 We determine these  parameters using  the protein structures in  \cite{select}. We use
the list of proteins in http://bioinfo.tg.fh-giessen.de/pdbselect  dated February 11, 2011.   The structures in this list have a
resolution better than 3.0 \.A, R-factor  less than 0.3, and less than $25\%$ homology equivalence.  But  since
our ambition is 
to match B-factor accuracy of $0.6-0.7$ \.A  we have further pruned this list  by selecting only those x-ray structures that have resolution better 
than 2.0 \.A. This leaves us with a total of 3.027 proteins.  With a very few exceptions the R-factors in our pruned set  
are  less  than 0.25, and the mean value is R=0.17, see Figure 1.
\begin{figure}[h]
\includegraphics[scale=0.13,clip=true]{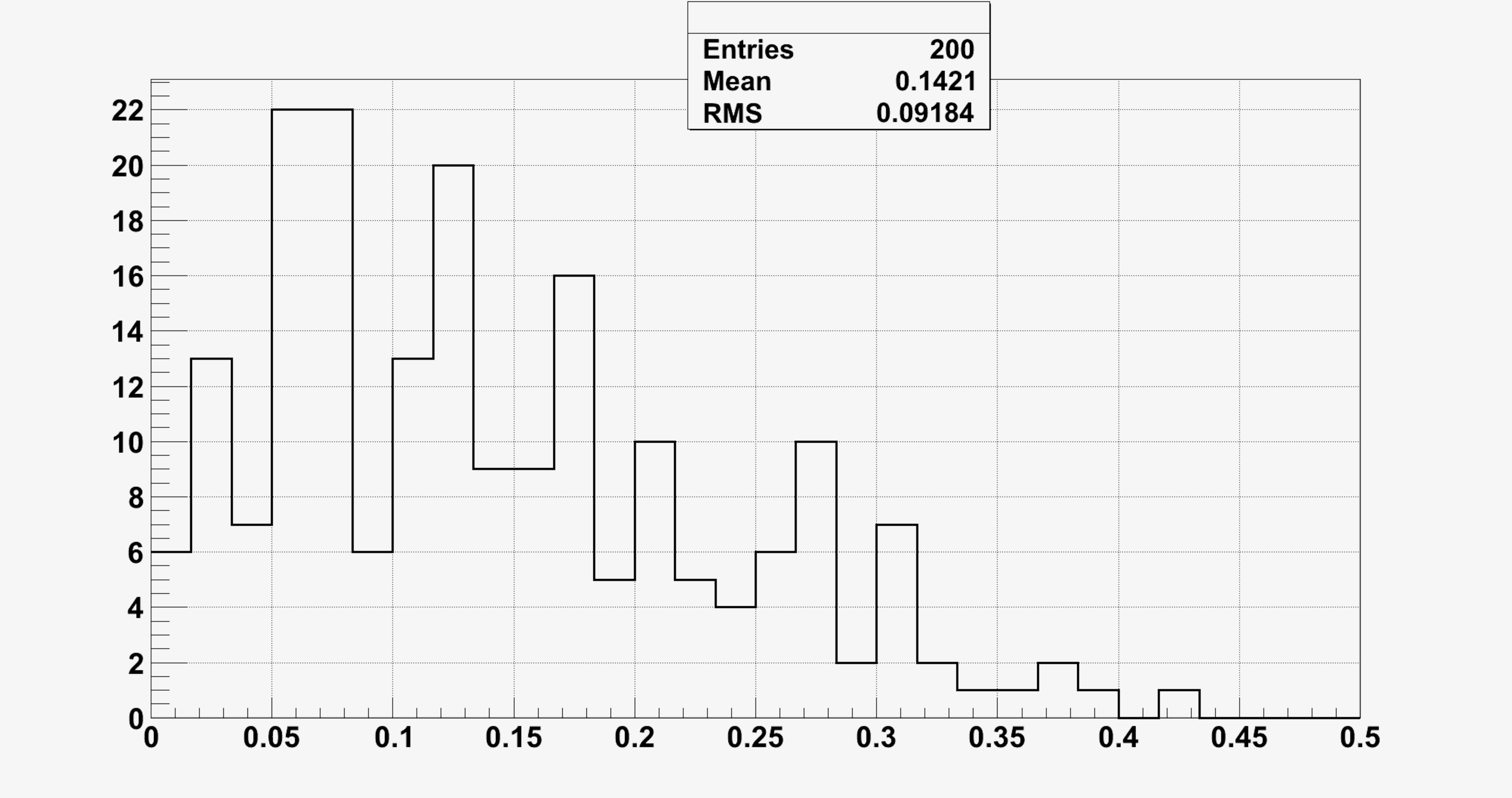}
\caption{The distribution of the RMSD distance between the original 200 loops and their soliton approximations. }
\label{fig:02}
\end{figure}
In Table I we display the distribution of the residues in our data set according to the different secondary structures.
{
\begin{table}[h]
\caption{The total number of residues in our data set and their breakdown into different structures according to PDB. 
} \vskip 0.5cm
\begin{tabular}{|c|c|c|c|}
\hline 
total & helices  & strands & loops  
\tabularnewline \hline
550.997  & 216.732  & 140.625 & 193.640
\tabularnewline\hline
\end{tabular} 
	\label{tab:para3a}
\end{table}
}

Our construction of the parameters in  (\ref{bond}), (\ref{Etau})
proceeds in three steps:  We first use visual inspection  and RMSD minimization to identify a set of 200 different putative fundamental
loop structures  that describe  the loops in our list of proteins  with  different pre-defined accuracies.  We then determine 
the parameters  (\ref{bond}), (\ref{Etau}) so that the ensuing profiles approximate our 200 visually identified loop 
structures with the RMSD precision of 0.5 \.A or better. 
Finally, we consider various multiple coverings of the fundamental domain $[-\pi, \pi]$
of the bond angle, to determine a set of integers $N_1$, $N_2$ in (\ref{bond}).  The aim is to shrink the set of 200 loop structures
into a smaller subset that covers the original set with an accuracy that exceed  0.5 \.A in RMSD distance.

We start our construction by selecting a  random 
protein from our list, for example  the myoglobin with PDB code 1A6M.  In Table II we present the loop structures  
that we have visually identified in 1A6M. For this we  have analyzed its ($\psi_i, \theta_i$) profile using  the symmetry transformation (\ref{dsgau}) 
in the manner we have  explained in \cite{cherno} (see also Figure 6b).
In addition, we list the number of  times each of the loops  
appears in our entire data set.  For this we identify two loops provided  they have the same length and their mutual
RMSD distance is less than 0.5\.A.
{
\begin{table}[h]
\caption{The sites of the loop structures (\ref{bond}), (\ref{Etau}) that we identify in 1A6M.  Indexing starts from the N terminus.
We also display the number of matches we have in our data set  when we use as a cut-off value 0.5  \.A in RMSD distance. 
} \vskip 0.5cm
\begin{tabular}{|c|c|}
\hline 
Sites &1Matches
\tabularnewline \hline
8-23 &  525
\tabularnewline \hline
34-39 & 702
\tabularnewline \hline
41-46 & 610
\tabularnewline \hline
48-54 & 183  
\tabularnewline \hline
56-61 & 819
\tabularnewline \hline
77-83 & 2
\tabularnewline \hline
81-87 & 1501
\tabularnewline \hline
95-100 &  298
\tabularnewline \hline
96-100 & 2352
\tabularnewline \hline
122-127 & 287 
\tabularnewline \hline
\end{tabular} 
	\label{tab:para3}
\end{table}
}

\noindent
In  the sites for the loop structures that we list  in Table II,  the first and last sites  always coincide with 
values that describe known regular secondary structures such as  (\ref{rhab}). 
Consequently for example the loop 18-23 has four sites in the loop proper, and the first and last sites 18 and 23 are
in $\alpha$-helical positions as far as the parameter values are concerned.

It is notable that two pairs of putative loops, the loops (77,83) and (81,87) and in 
particular the loops  (95,100) and (96,100) are overlapping.  
In the latter case this is because  
we can  introduce  two different interpretations: We can either
interpret (95,100)  as a loop that connects an $\alpha$-helix  with another $\alpha$-helix, while by removing the site 95 
we have a configuration that we can interpret as a loop that starts from a $\beta$-strand. 
A refinement of the cut-off RMSD distance 0.5 \.A to a smaller value might help us to
eliminate one of these two loops. However, this would be highly questionable as  it would also push us 
below the experimental $B$-factor accuracy and that does not make much sense. 
We adopt the position that 0.5 \.A is about the best one can do in identifying the fundamental loops, 
with presently available experimental data.

We continue  by selecting a new protein structure. 
We perform the same visual identification of loops. 
We continue the process until we have identified a total
of exactly 200 loops such  that {\it each pair} of these loops, with the same number of sites, 
has a mutual RMSD distance that  exceeds  0.5 \.A.  
For this we only need to go thru 44 randomly chosen protein structures in our data set, the proteins
are listed in  Table III. 
{
\begin{table}[h]
\begin{tabular}{|c|c|c|c|c|}
\hline 
1A6M(A)  & 2OVG(A)  & 2O7A(A)  & 1XG(D)  & 1LWB(A) 
\tabularnewline \hline
1SAU(A) & 2I4A(A) & 3GOE & 2AIB & 1P6O 
\tabularnewline\hline
2VZC(A) & 1WMA(A) & 3F1L(B) & 1MUN(A) & 3PD7(B) 
\tabularnewline\hline
1WKQ(B) & 3E7R(L) & 3OQ2(A) & 3BFQ(G) & 1SEN(A) 
\tabularnewline \hline
1MN8(C) & 3CT6(A) & 2XL6(A) & 3A5F(B) & 3CI3(A) 
\tabularnewline \hline
3G46(A) & 1ZZK(A) & 1PSR(A) & 1I27(A) & 1P1X(B) 
\tabularnewline \hline
2V9V(A) & 2W72(A) & 1OAI(A) & 3DNJ(A) & 1NNF(A) 
\tabularnewline \hline
3LB2(A) & 1Q6O(B) & 3P3C(A) & 1QNR(A) & 3L0F(A) 
\tabularnewline \hline
1DO4(A) & 3OGN(B) & 3MBX(B) & 2W91(A)  & -
\tabularnewline \hline
\end{tabular} 
	\label{tab:1}
	\caption{The PDB codes of the 44 proteins that we have used in constructing our loop library  (with chain in parentheses)} 
\end{table}
}

These 200 loop structures have between 5 and 9  sites, including the two end points that are in regular secondary structure
positions.  The distribution of the number of loops according to their size is shown in Table IV.  
{
\begin{table}[h]
\caption{The distribution of the 200 loop structures according to their length, 
with the first and last sites  in regular 
secondary structure positions. Two loops with the same length but separated from each
other by more than 0.5 \.A in RMSD distance are considered different. } \vskip 0.5cm
\begin{tabular}{|c|c|c|c|c|c|}
\hline 
Length & 5 & 6 & 7 & 8 &9
\tabularnewline \hline
Number & 32 & 116 & 44 & ~7~ & ~1~ 
\tabularnewline\hline
\end{tabular} 
	\label{tab:para3b}
\end{table}
}

Loops with 
length 6 are by far the most common and
we only identify seven length 8 loops, only one fundamental loop with length 9, and none longer.  We suspect 
that the  very few length 8 loops and the single length 9  loop can probably be interpreted as combinations 
of  length 5 and 6 loops by an extended search of fundamental loops; The purpose of the present article is not to 
develop a publicly available databank but to form a conceptual basis for developing such a databank by showing that we have a method that works. Consequently we have 
stopped our search of new loop structures when we reached exactly  200 structures.

In Table V we display how many residues in our  entire data set are covered  
by our 200 loops,  when we search for structures using as a criterion the RMSD distance between the structure
and a loop.  
We have performed the search with RMSD cut-off values that range  from  0.2 \.A to  0.7 \.A.  
The largest  value 0.7 \.A is selected to slightly exceed the estimate (\ref{x}).
{
\begin{table}[h]
\caption{The coverage of our putative loops in terms of residues, at different RMSD cut-off values. Note that  a structure that has  
between 5 to  9 sites, has a length that is roughly between  20-40  \.A.
} \vskip 0.5cm
\begin{tabular}{|c|c|}
\hline 
RMSD cut-off (\.A) & Loop sites matched
\tabularnewline \hline
 $<$ 0.2 & 7.208
\tabularnewline \hline 
 $<$ 0.3 & 31.655
 \tabularnewline \hline
  $<$ 0.4 & 78.561
\tabularnewline \hline  
  $<$ 0.5 & 148.267
\tabularnewline \hline  
  $<$ 0.6 & 245.954
\tabularnewline \hline  
  $<$ 0.7 & 428.387
\tabularnewline \hline
\end{tabular} 
	\label{tab:para4}
\end{table}
}
\noindent
For a cut-off value of 0.6 \.A {\it i.e.} just below (\ref{x}) the number of sites  in configurations that are covered  
by our 200 loops  already clearly exceeds the total number of  sites that are classified as loop sites according to PDB; see
Table I.  This suggests that we cover all of the loop structures.  However, a closer 
inspection shows that due to overlapping structures
the actual coverage is somewhere  around 90$\%$. But when the cut-off value reaches 0.7 \.A we rarely find any 
loop structures that remain uncovered. Since we have  a very representative data set, this
proposes that within the experimental 
B-factor fluctuation distance accuracy (\ref{x}), a large majority of {\it all} loops in PDB, both short and long,  are various kind 
of modular combinations of the  200 {\it fundamental} loops we have identified.

We now proceed to the second step of our construction. Here we search for parameters in the soliton profile 
(\ref{bond}), (\ref{Etau}) that describe  
our fundamental 200 loops, so that the RMSD distance between a loop and its soliton is less than 0.5 \.A.
Since the RMSD distance between any two loop structures  
in our set of 200 loops is always larger than 0.5 \.A,  
we demand that the pairwise RMSD distance between any two explicit solitons also exceeds 0.5 \.A.    
We estimate the parameters using a Monte Carlo search that minimizes the RMSD
distance between a loop and its soliton.  
The parameter values are summarized in Figures 2-4:  
\begin{figure}[h]
\includegraphics[scale=0.15,clip=true]{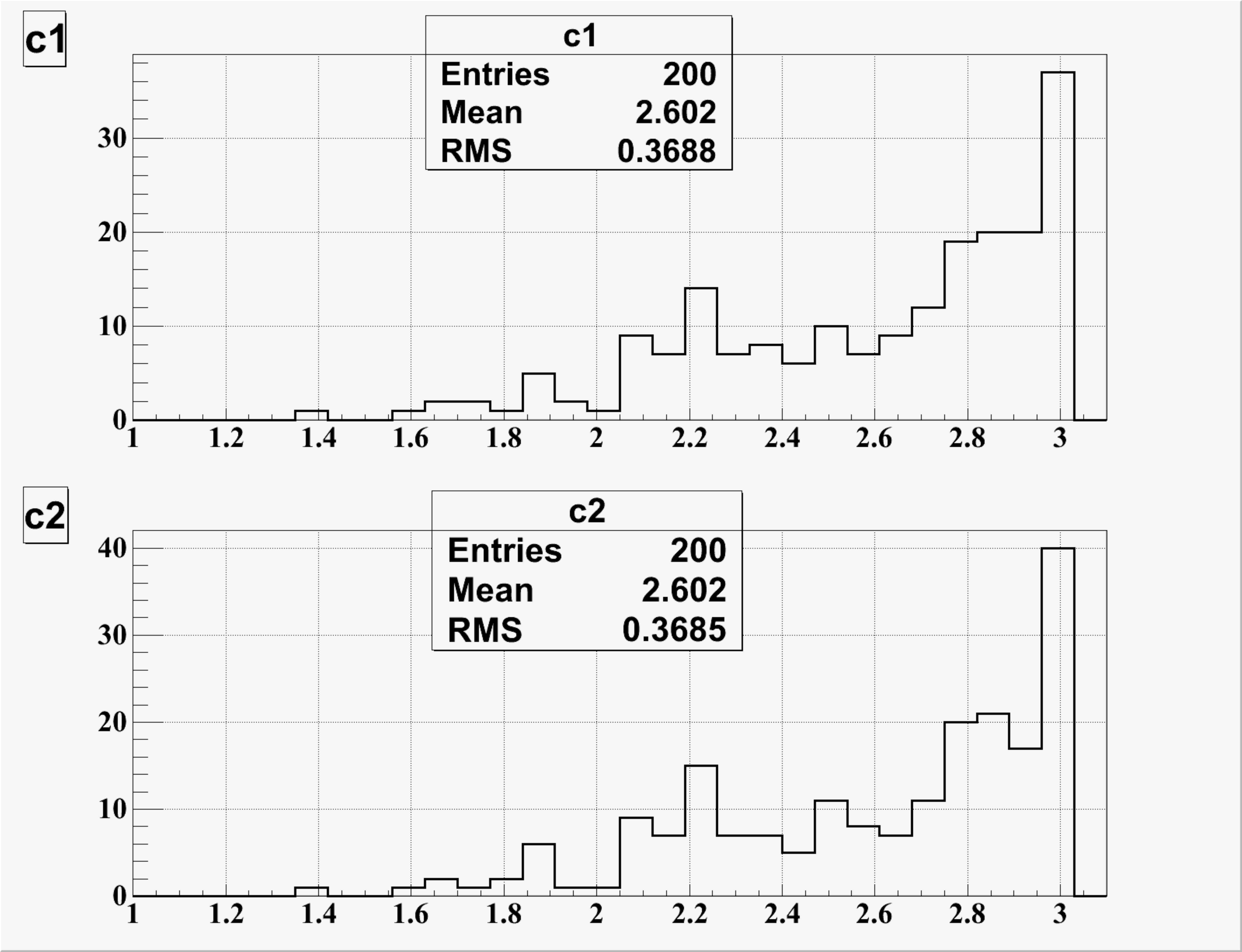}
\caption{The distribution of of the parameter value $c_1$ and $c_2$ in  (5) in the 200 solitons we have constructed. As expected,
these two distributions are practically identical. }\label{fig:03}
\end{figure}

\begin{figure}[h]
\includegraphics[scale=0.15,clip=true]{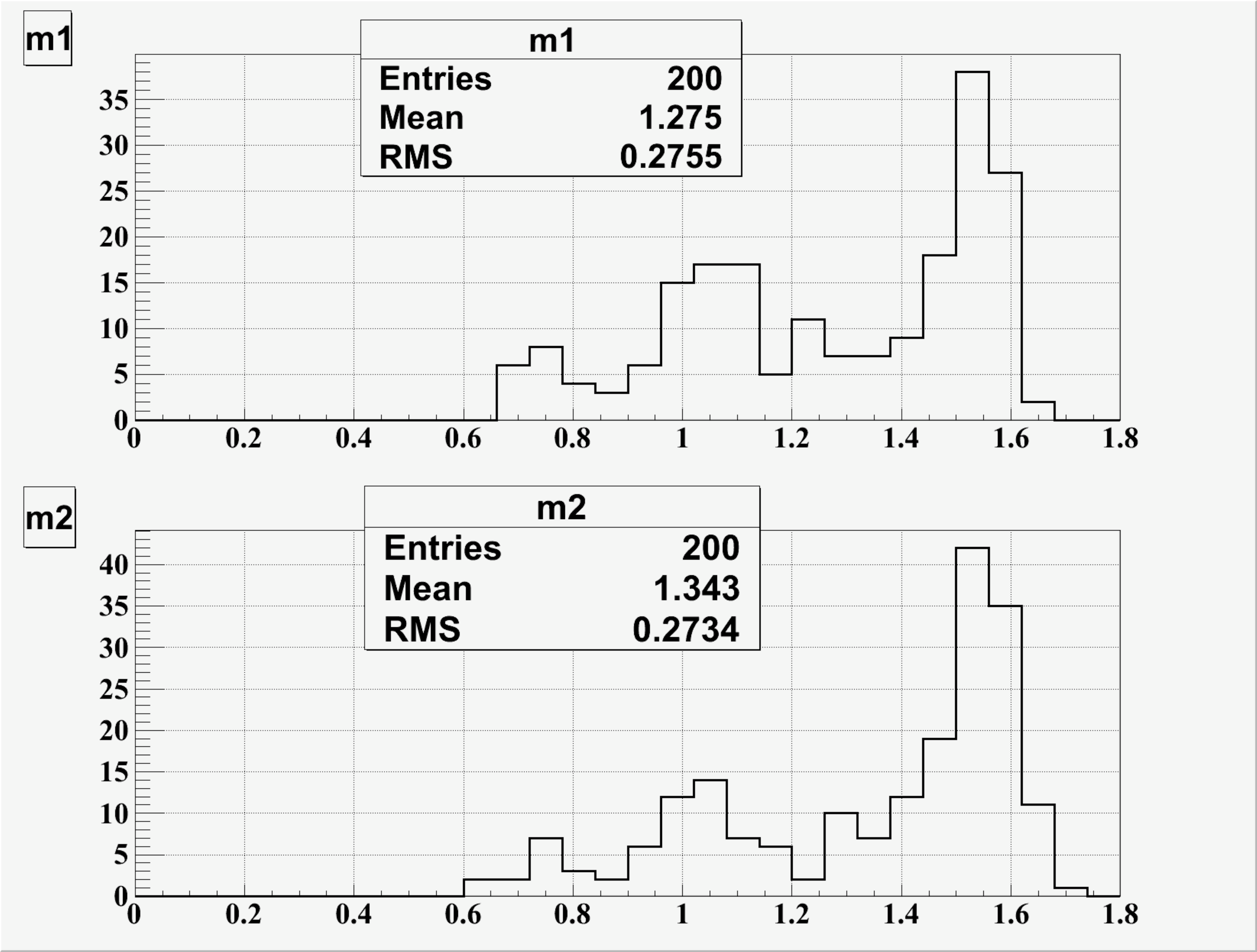}
\caption{The distribution of of the parameter values $m_1$ and $m_2$  in  (5)  for  the 200 solitons.  
As in Figure 1, the distributions are highly symmetric, the difference is not statistically meaningful. 
The $\alpha$-helices  and $\beta$-strands (3) are also 
clearly identifiable in the parameter values.
}
\label{fig:4}
\end{figure}

\begin{figure}[h]
\includegraphics[scale=0.14,clip=true]{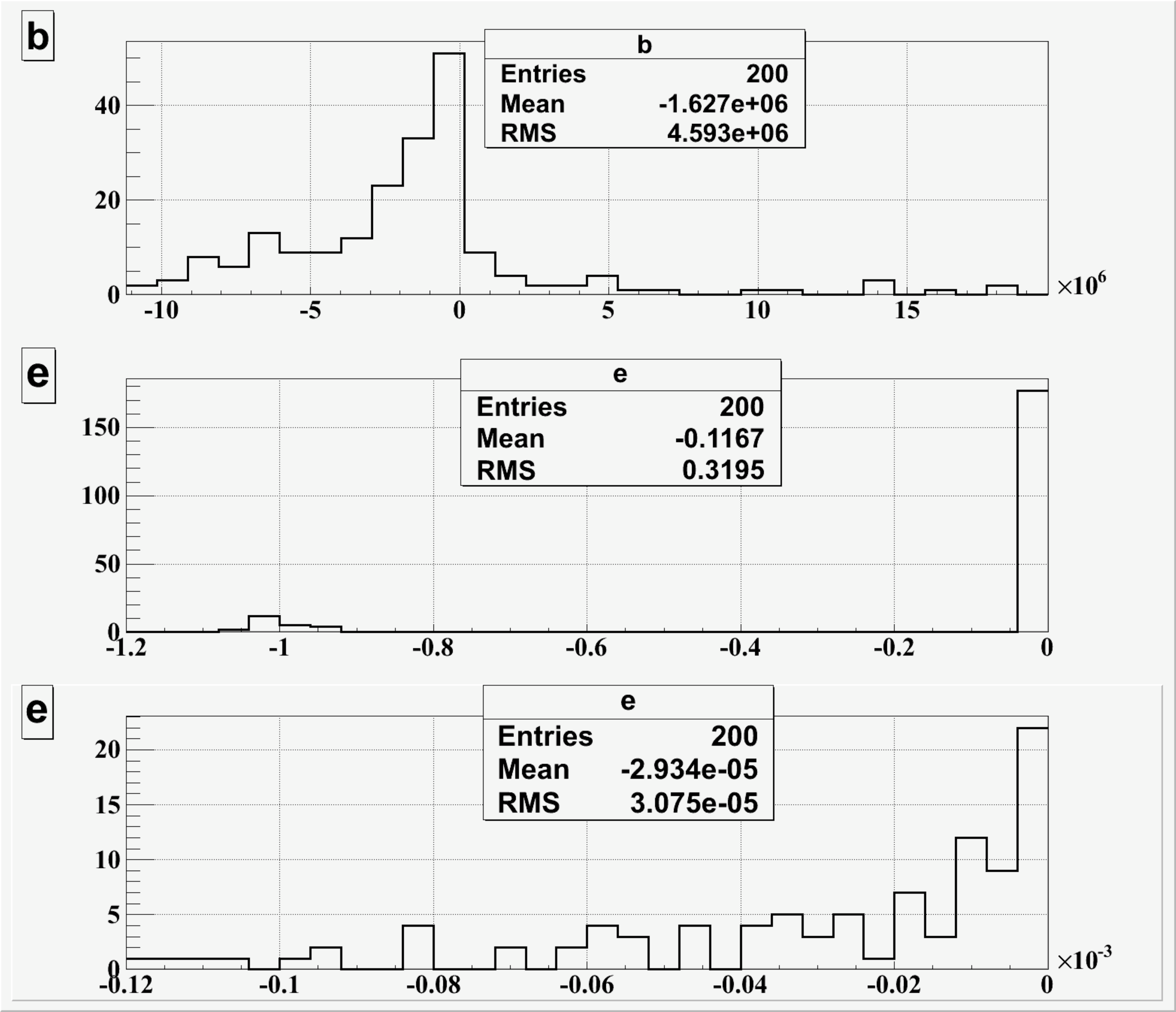}
\caption{The distribution of of the parameter values for the torsion angle (6) in our 200 solitons. Observe that the parameter $e$ clusters
in two regions, around -1 and below -10$^{-4}$. Furthermore,
the parameter $b$ has very large values, in excess of $\pm$10$^{6}$ and the spread is very large. The fundamental region of the torsion
angle $\theta_i$ is $[-\pi, \pi]$ and the large values reveal that as a soliton, the loops cover the spheres $(\psi,\theta) \sim \mathbb S^2$
several times {\it i.e.} each of the loop is a multiple soliton configuration. This explains why a very regular 
soliton such as (5), (6)  can model the apparently highly irregular $\psi_i$ and $\theta_i$ profiles such as in
that we commonly find in PDB.
}
\label{fig:05}
\end{figure}

For each of
the 200 loops, we are able to identify parameters so that there is always a soliton profile (\ref{bond}), (\ref{Etau}) with explicit
parameter values, that describes the
loop with RMSD accuracy that is less than  0.5 \.A.   In fact,  as shown in Figure 1  
the mean RMSD distance between the original loop
configuration and its explicit 
soliton is a mere 0.14 \.A, slightly less than the 0.15 \.A estimate for zero point fluctuations in \cite{andrei}.
At this separation distance, it then becomes conceptually meaningless to consider the two structures as different.

In Table VI we show how the number of sites that our solitons cover in our full data set depends on the cut-off RMSD distance, 
for values between 0.2 and 0.7 \.A.
The results are very similar to those in Table V, there is no practical difference. We also find that when the RMSD cut-off value
exceeds (\ref{x}), the loop structures in our data set that are not fully covered by our 200 explicit solitons become very rare. Consequently we have
 succeeded in constructing a basis of 200 explicit soliton structures that cover most of the PDB loops, apparently over 90\% of them, 
 and with an accuracy
 that is comparable to the experimental B-factor accuracy.
{
\begin{table}[h]
\caption{The coverage of our  explicit soliton configurations in terms of residues, at different RMSD cut-off values.
} \vskip 0.5cm
\begin{tabular}{|c|c|}
\hline 
Cut-off (\.A)&Loop sites matched
\tabularnewline \hline
 $<$  0.2 & 5.954 
 \tabularnewline \hline
 $<$ 0.3 & 28.399 
 \tabularnewline \hline
 $<$ 0.4 & 74.037 
 \tabularnewline \hline
 $<$ 0.5 &  144.683
 \tabularnewline \hline
 $<$ 0.6 & 245.257
 \tabularnewline \hline
 $<$ 0.7 & 433.737
\tabularnewline \hline
\end{tabular} 
	\label{tab:para6}
\end{table}
}

Finally, we have constructed our 200 explicit 
solitons by a direct approach, with no attempt for optimization. As a consequence we
suspect that the number of explicit solitons  can be substantially decreased without compromising the
coverage.
To show that this is the case we have employed the freedom to choose the integers $N_1$ and $N_2$ in (\ref{bond}) 
independently. These integers are covering numbers, they determine how 
many times we cover the  fundamental domain $\psi \in [-\pi, \pi]$. 
They have no effect how the parameters $m_1$ and $m_2$ determine
the asymptotic $\psi_i $ values.
Consequently two solitons that differ from each other only by these integers 
interpolate between regular secondary structures  with identical $\psi$ values, and  in this sense 
they can be viewed as different multiple coverings of a single basic soliton with $N_1 = N_2 = 0$.
But note that the $\theta_i$ values can still be different.

We proceed as follows: We first select a pair of solitons in our library.  
All the parameters in the first soliton are kept fixed.  In the second soliton we  also keep all 
parameters fixed, except that we  allow the integers $N_1$ and $N_2$ to vary.  We then ask 
whether it is possible to find a new set  of integers ($N_1, N_2$) in the second soliton, so that the RMSD 
distance between the two solitons becomes less than 0.5 \.A. 
We have found that it is possible to substantially lower the RMSD distance between two solitons. 
For example, one can find pairs where the initial distance is above 3.5 \.A  and this becomes lowered to  a mere
0.28 \.A when we judiciously select the
integers ($N_1, N_2$).  In this way we have been able to show that in our set of 200 solitons there are only  57 covering solitons 
that we fail to bring to within a distance of 0.5 \.A from each other.  But we suspect that in a carefully constructed and optimized  
library the number  of covering solitons is  even much smaller. 

\section{Examples}

As an example how the 200 explicit 
solitons cover our data set at different cut-off values, we show  in Figure 5 the typical $\psi_i$ profile of 
a protein in PDB, we have randomly chosen the one with PDB code 1KZQ. This protein has 289 residues, the experimental 
resolution  is 1.7\.A and  the observed R-value is 0.2.
In the top Figure 5
we use the cut-off value 0.3 \.A to locate our solitons. This cut-off value is clearly below the B-factor accuracy 
of the C$_\alpha$ atoms in 1KZQ,  and our 200 solitons cover only around 20 per cent of the loop structures. This coverage is 
consistent  with results in Table VI.  When we increase the cut-off value to 0.5 \.A (middle Figure 5) 
most of the loops become covered by solitons, and at 0.7\.A there is only one loop with three sites within the loop ({\it i.e.}
a soliton with 5 sites),
that does not appear among our 200 solitons. 
This loop can be modeled by a single soliton and the soliton can be added to our initial unpruned library if so desired, increasing the number of 
solitons to 201.  Alternatively, we could try to
describe it as a multiple-covering of one of our 57 solitons. Notice that in addition there are four 
isolated sites where the deviation exceeds the cut-off value of 0.7 \.A.
Indeed, it is not too exceptional for proteins that are resolved with this resolution to have individual C$_\alpha$ 
sites where the experimental accuracy as measured by the B-factor fluctuation distance exceeds 0.7 \.A.  These low-resolution C$_\alpha$
carbons commonly become visible in our matching procedure, and this could be used to identify potential problems in data.
\begin{figure}[h] 
\includegraphics[scale=0.095,clip=true]{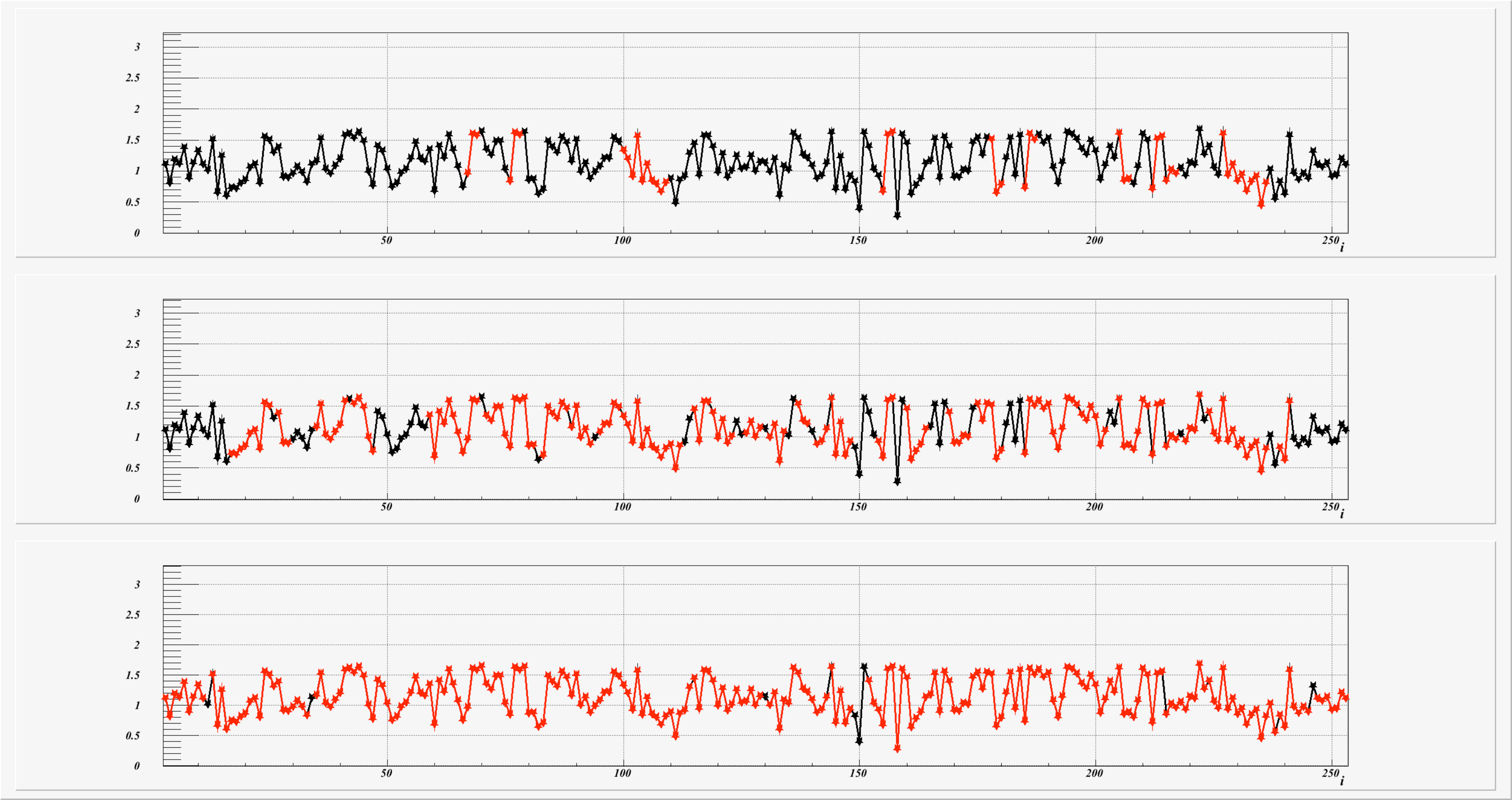}
\caption{An example how our 200 explicit solitons cover the protein
with PDB code 1KZQ in our data set, in terms of the bond angles $\psi_i$. 
We use cut-off values 0.3 \.A (top), 0.5 \.A  (middle) and 0.7 \.A (bottom). Red dots and lines correspond to sites and structures that are described by the solitons with the cut-off accuracy or better, while black 
dots and lines correspond to sites where the local distance exceeds the cut-off value;  isolated black 
dots indicate local fluctuations in B-factors.  Three or more consecutive black dots indicate the presence of a 
loop that is not covered by our 200 solitons. Note that at resolution 0.7 \.A (bottom) there is only one such loop.}
\end{figure}

As a second example we discuss  a loop in the protein with PDB code 3DLK.  
In \cite{andrei} we showed how to construct  a soliton that describes 
the super-secondary structure that is located between the coordinate sites 398-416 in the A chain of 3DLK,
with RMSD accuracy 1.13 \.A.  The structure describes a loop that connects an $\alpha$-helix to 
a $\beta$-strand.  We now analyze this  loop in terms of our library of 200 solitons.  
\begin{figure}[h]
\includegraphics[scale=0.1,clip=true]{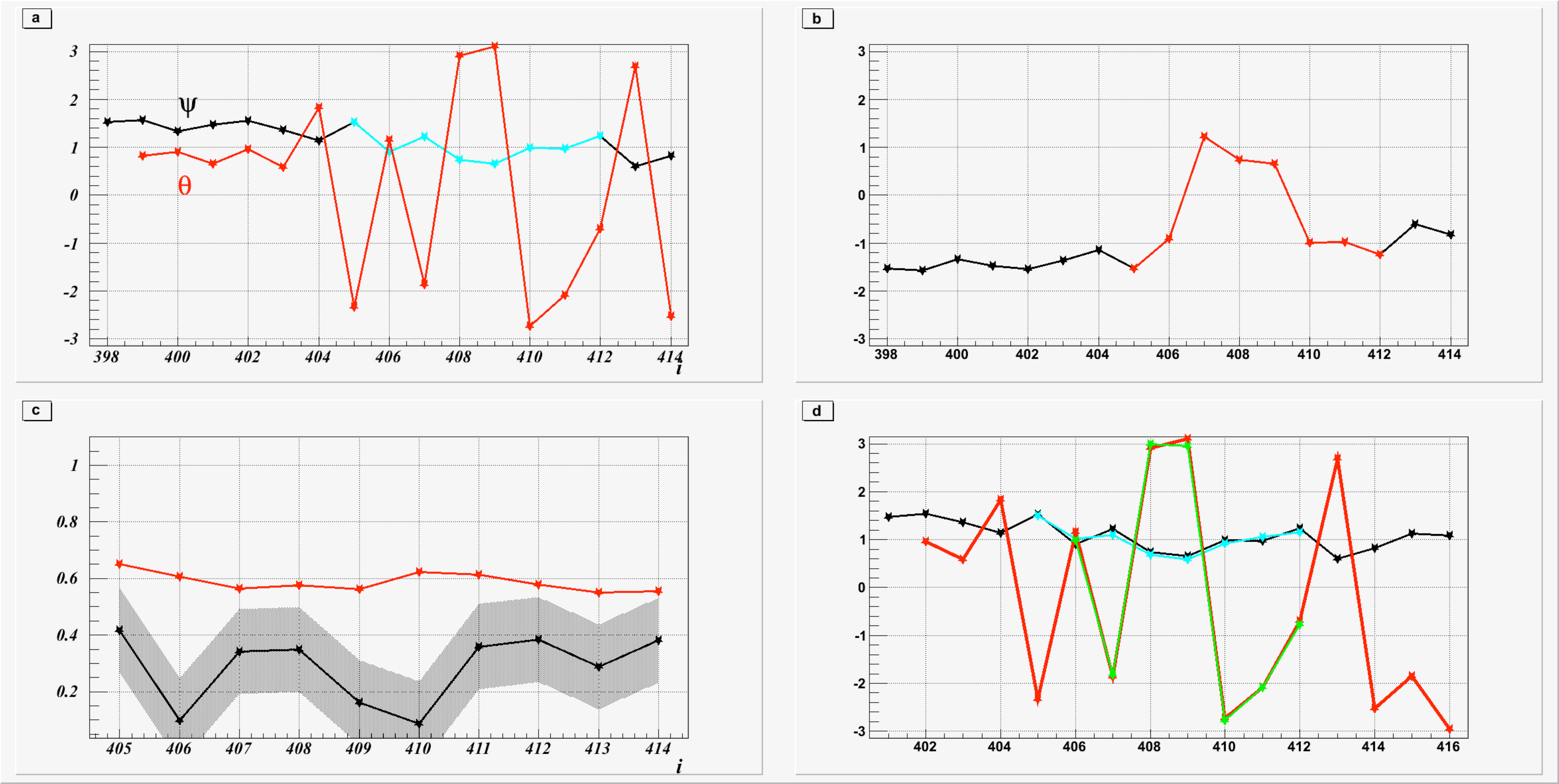}
\caption{Figure a) shows the PDB data for the bond and torsion angles for the monomer A in 3DLK, for sites 398-414 
Figure b) displays the $\psi_i$ profile, after we have introduced the gauge transformation (\ref{dsgau}). The two solitons 
are clearly visible between sites 405-412. In Figure c) we compare the B-factor of the 3DLK (upper line in red) 
with the distance between its  backbone and the  two-soliton configuration (lower line with black). The shaded  area is the 0.15 \.A
fluctuation regime around the soliton. In Figure d) we compare the ($\psi_i, \theta_i$) distributions for the PDB data (black and red)
and the two-soliton configuration (blue and green).}
\label{fig:04}
\end{figure}
In Figure 6a we display the $(\psi_i, \theta_i)$ profile around the loop region; We remind that according to (\ref{tb}), (\ref{bt})
the $\psi_i$ is determined by the three coordinate sites $i$, ${i+1}$ and ${i+2}$ while  $\theta_i$ is determined
by the four sites with indices from $i-1$ to $i+2$. 

There is a relatively  large local fluctuation at the coordinate site $i$=404, according to the PDB data the B-factor 
of the C$_\alpha$ atom at this site is 40.0 (\.A$^2$) which is clearly above (\ref{bmax}). 
The B-factors at the coordinate sites 403 and 405 are also relatively high, with values  33.5 and 33.5  respectively.  But
beyond the coordinate site  405, the B-factors are around 25-30  that is the  Debye-Waller fluctuation distances are below 0.7 \.A 
for the sites that we have displayed in Figure 6. In Figure 6c  the top  (red) line shows the fluctuation distances for the
coordinate sites 405-414.

In Figure 6b we display the profile of $\psi_i$, after we have  implemented the transformation (\ref{dsgau}).
We clearly identify  two soliton profiles (\ref{bond}). Due to the relatively large  B-factor at coordinate site 404, 
we try and take the first soliton to start from the bond angle site 405.  The definition of this bond angle is independent of
site 404, and thus we are optimistic that we do not need to compromise with our ambition to exceed the B-factor accuracy (\ref{bmax})
in our loop description. The second soliton ends at bond angle site 412, and the  two solitons  overlap 
between the bond angle sites 407 and 409.  
When we search for similar structures among our 200 solitons, we find two profiles that in combination 
match the loop. The first soliton covers the coordinate
sites 405-410,  and the second soliton covers  the coordinate sites 409-414. In terms of the bond angles, together they cover the sites
405-412. When we combine these two solitons so that they match each other as accurately as possible
at their common coordinate sites 409 and 410, we find a two-soliton configuration
that describes the protein loop  for residues 405-414 with a RMSD accuracy of 0.31 \.A. The (lower) black  line in Figure 6c shows the difference
between the PDB structure and the two-soliton structure. This difference is clearly less than the Debye-Waller 
B-factor distance, at every site. The shaded area
describes the zero point fluctuation regime around the solitons. We have followed \cite{andrei} to estimate 
that the zero point fluctuations have an amplitude that is no larger than 0.15 \.A.  Finally, in Figure 6d we 
compare the $\psi_i$ and $\theta_i$ values of the PDB data and the two-soliton configuration. There is essentially 
no difference.

\section{Conclusion}

Protein loops  remain a major challenge both in structure classification and prediction.
Loops are commonly viewed as apparently random regions  with no regular self-similar  
structure.  Here we have shown that loops are not random at all.
Their shape is fully determined and with experimental B-factor accuracy  by the dark soliton solution of
a generalized discrete nonlinear Schr\"odinger equation that has
only  two loop specific parameters. In particular we have found that the number of  different parameter sets {\it i.e.}
fundamental loops appears to be no more  than 200  and probably it is even smaller than 57 if we allows for multiple coverings.
When the fundamental loops together with the helices and strands are at our disposal, the construction of entire 
folded proteins becomes  like a play with Lego bricks. We can build the  entire
protein from these modular components by simply putting  them together, one after another. 
Moreover, our  quantitative approach  is firmly grounded on a Physics 
based energy function. This should  enable  energetic analyses of protein folding, 
and energy comparisons between folds and misfolds. We propose that our soliton approach to protein folding
can add a powerful  component to the existing classification and modeling schemes.

\end{document}